\documentclass[10pt]{article}
\usepackage{graphicx}
\usepackage{subfigure}
\usepackage{amsmath, amssymb}
\usepackage{multirow} 
\usepackage{hhline}

\newcommand{\lsim}{\lesssim}

\newcommand{\beq}{\begin{equation}}
\newcommand{\eeq}{\end{equation}}
\newcommand{\bea}{\begin{eqnarray}}
\newcommand{\eea}{\end{eqnarray}}

\newcommand{\fb}{{\rm fb}}
\newcommand{\br}{{\rm BR}}
\newcommand{\tev}{{\rm TeV}}
\newcommand{\gev}{{\rm GeV}}

\newcommand{\met}{{/\!\!\!\! E_T}}

\def\Title#1{\begin{center} {\Large #1 } \end{center}}
\def\Author#1{\begin{center}{ \sc #1} \end{center}}
\def\Address#1{\begin{center}{ \it #1} \end{center}}

\newcommand\pubblock{\rightline{\begin{tabular}{l} Proceedings of the Second Annual LHCP\\ \pubnumber\\
         \pubdate  \end{tabular}}}

\newenvironment{Abstract}{\begin{quotation} \begin{center} 
             \large ABSTRACT \end{center}\bigskip 
      \begin{center}\begin{large}}{\end{large}\end{center} \end{quotation}}

\newenvironment{Presented}{\begin{quotation} \begin{center} 
             PRESENTED AT\end{center}\bigskip 
      \begin{center}\begin{large}}{\end{large}\end{center} \end{quotation}}

\def\Acknowledgements{\bigskip  \bigskip \begin{center} \begin{large}
             \bf ACKNOWLEDGEMENTS \end{large}\end{center}}

\def\beq{\begin{equation}}
\def\eeq#1{\label{#1}\end{equation}}
\def\eeqn{\end{equation}}

\def\beqa{\begin{eqnarray}}
\def\eeqa#1{\label{#1}\end{eqnarray}}
\def\eeqan{\end{eqnarray}}

\let\bar=\overbar

\def\Dslash{\not{\hbox{\kern-4pt $D$}}}
\def\dslash{\not{\hbox{\kern-2pt $\del$}}}

\def\msb{{\bar{\ssstyle M \kern -1pt S}}}

\usepackage{color}

 \newcommand\pubnumber{ }

\newcommand\pubdate{\today}

\begin{document}

\large
\begin{titlepage}
\pubblock

\vfill
\Title{Charged Higgs Probes of Dark Bosons at the LHC}
\vfill

\Author{Kyoungchul Kong}
\Address{Department of Physics and Astronomy, University of Kansas, Lawrence, KS 66045, USA}
\Author{Hye-Sung Lee}
\Address{Department of Physics, College of William and Mary, Williamsburg, VA 23187, USA}
\Address{Theory Center, Jefferson Lab, Newport News, VA 23606, USA}
\Author{Myeonghun Park}
\Address{Kavli IPMU (WPI), The University of Tokyo, Kashiwa 277-8583, Japan}
\vfill
\begin{Abstract}
A very light (GeV scale) dark gauge boson ($Z'$) is a recently highlighted hypothetical particle that can address some astrophysical anomalies as well as the $3.6 \sigma$ deviation in the muon $g$-$2$ measurement.
We suggest top quark decays as a venue to search for light dark force carriers at the LHC.
Such $Z'$s can be easily boosted, and they can decay into highly collimated leptons (lepton-jet) with large branching ratio.
We investigate a scenario where a top quark decays to $b W$ accompanied by one or multiple dark force carriers and find that 
such a scenario could be easily probed at the early stage of LHC Run 2.
\end{Abstract}
\vfill

\begin{Presented}
The Second Annual Conference\\
 on Large Hadron Collider Physics \\
Columbia University, New York, U.S.A \\ 
June 2-7, 2014
\end{Presented}
\vfill
\end{titlepage}
\def\thefootnote{\fnsymbol{footnote}}
\setcounter{footnote}{0}
%

\normalsize 


\section{Introduction}

A very light vector boson (such as dark photon and dark $Z$) has attracted great attention because of the various astrophysical anomalies, $g_\mu$-$2$ anomaly, etc. With such appealing motivations, there are many active searches for light dark force carriers \cite{Essig:2013lka}. 
Dark force is roughly of GeV scale and can be searched for at both low energy and high energy experiments. 
Major search strategies are based on bremsstrahlung at fixed target experiments or meson decays. 
Collider signals for certain SUSY models and decays of a Higgs boson into $Z^\prime$ have been studied as well.
In this article\footnote{This presentation is mainly based on our paper in Ref.~\cite{Kong:2014jwa}.}, we present a novel channel to look for the dark force using top quark decays at the LHC.
In particular, we consider a top decay to $b$ quark and a charged Higgs ($H^\pm$), which in turn decays to a $W$ and a dark force carrier \cite{Kong:2014jwa,Davoudiasl:2014mqa}.
We find that decay products of such a light dark $Z$ are highly collimated and a dedicated analysis would help in finding a new boson with the $t\bar t$ samples, which is, otherwise, hidden in huge backgrounds.

\section{\boldmath Dark $Z$ production via top quark}

Our study is based on two Higgs Doublet Model (2HDM) of Type I with a light $Z^\prime$ (dark $Z$) \cite{Davoudiasl:2012ag}.  
The dark $Z$ is a gauge boson of a new dark $U(1)$ with very weak couplings to the SM particles.
It has no direct couplings to the SM particles yet couples to them through the gauge kinetic mixing.
In particular, the $Z^\prime$ that we are interested in may have both axial and vector coupling via extra mass mixing.
Exact couplings depend on details of model, especially on how the $Z^\prime$ gets a mass, therefore dependence lies on details of Higgs sector.
The FCNC constraints in the 2HDM can be addressed by the new $U(1)$, under which Higgs doublets carry different charges.
The $Z'$ search in the SM-like Higgs decay was studied in Refs.~\cite{Davoudiasl:2012ag,Davoudiasl:2013aya}, but the existence of a charged Higgs in the model allows a new search mode.
The charged Higgs in this model can be much lighter than typical bounds as it can dominantly decay into $Z'$ final states as $W Z'$ \cite{Davoudiasl:2014mqa} or $W Z' Z'$ \cite{Lee:2013fda} depending on the masses of the non SM-like scalars.
In general, for a light $Z^\prime$, branching fraction into leptons is large.

If the dominant decay of the $Z'$ is into invisible light particles, the search will likely be more challenging and depend on how well the missing energy signal can be separated from the background \cite{Davoudiasl:2014mqa}. 
An approximate bound on this mode can be inferred from ATLAS/CMS bounds on stop production followed by stop decay to top + neutralino, 
which may constrain only a lower mass of the charged Higgs.
In this study, we will focus on the decay of $Z^\prime$ into dilepton. 
At the LHC, such a $Z^\prime$ can be boosted and two leptons from the $Z^\prime$ decay may appear as a lepton-jet.

Possible scenarios to look for the $Z^\prime$ in connection to the top quark decay are following.
\begin{enumerate}
\item $t \to b H^+ \to b W + Z'$,                          ~~~ through $H^\pm W^\mp Z'$ coupling
\item $t \to b H^+ \to b W + h \to b W + Z' Z'$,    ~~~ with a light non-SM Higgs boson $h$ 
\item $t \to b W^* \to b W + Z'$,                           ~~~ through $Z' W W$ coupling
\item $t \to b W^* \to b W + h \to b W + Z' Z'$,     ~~~ through $hWW$ coupling
\item $t \to q Z^\prime$ with $q=u, c$, ~~~ via the $W$ loop
\item $Z^\prime$ radiation off from the top quark
\end{enumerate}
For the rest of our study, we will focus on $t \to b H^+ \to b W^+ + Z'$ with $m_W \lsim m_{H^\pm} \lsim m_t$ \cite{Kong:2014jwa}.

\begin{figure}[t]
\begin{center}
\includegraphics[width=0.32\textwidth,height=0.23\textwidth,clip]{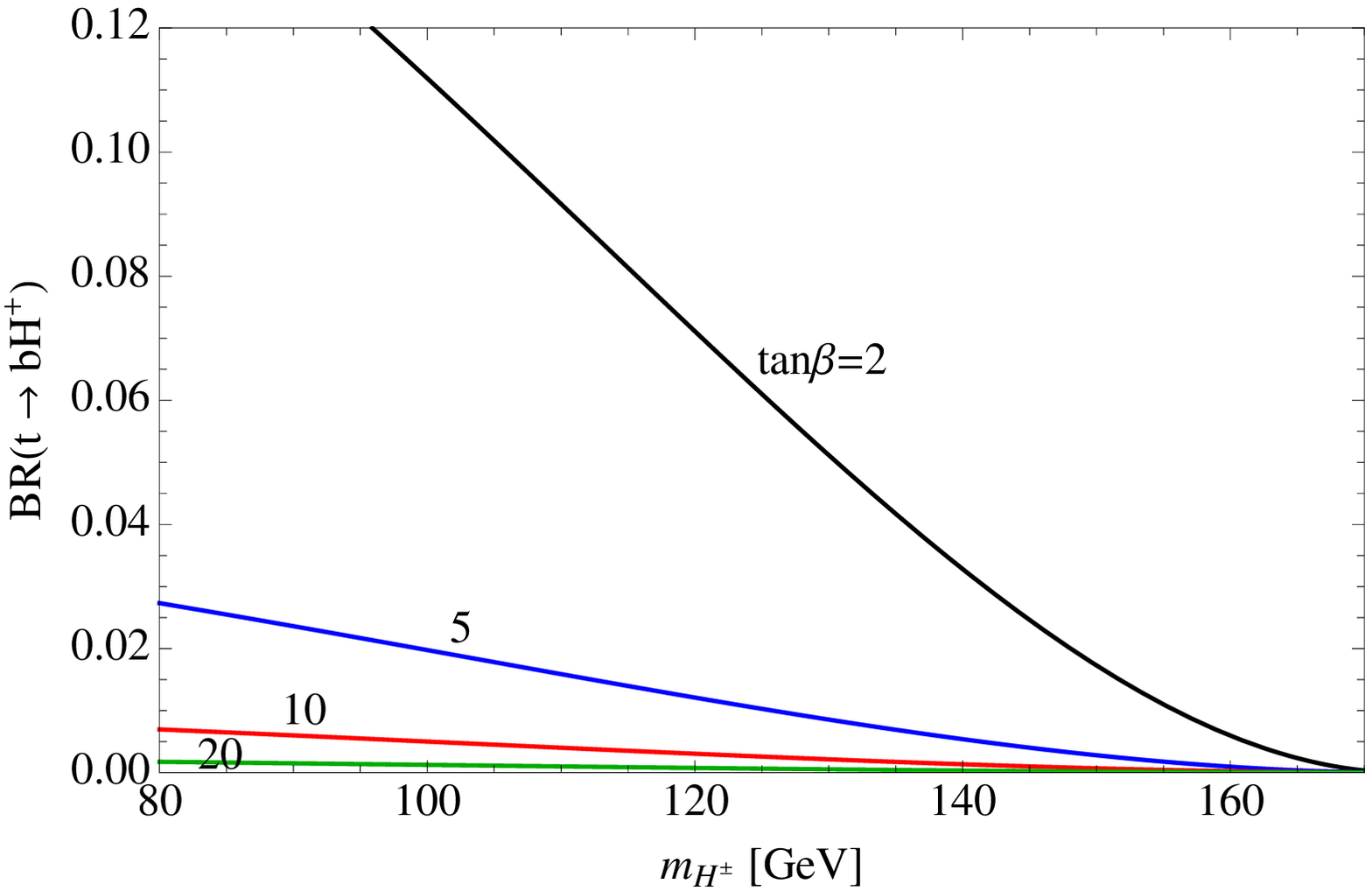} \hspace{0.1cm}
\includegraphics[width=0.32\textwidth,height=0.23\textwidth,clip]{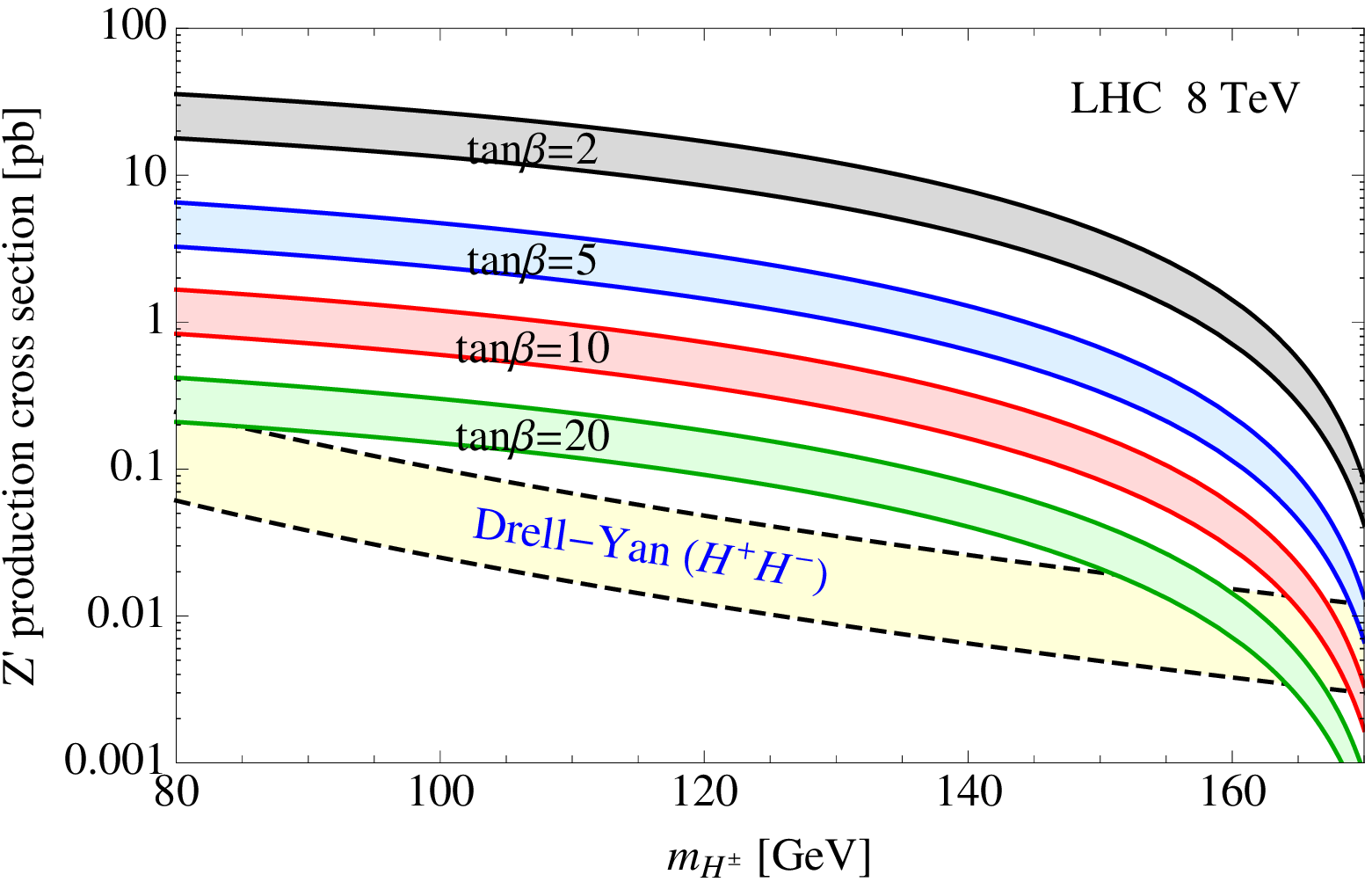} \hspace{0.1cm}
\includegraphics[width=0.32\textwidth,height=0.23\textwidth,clip]{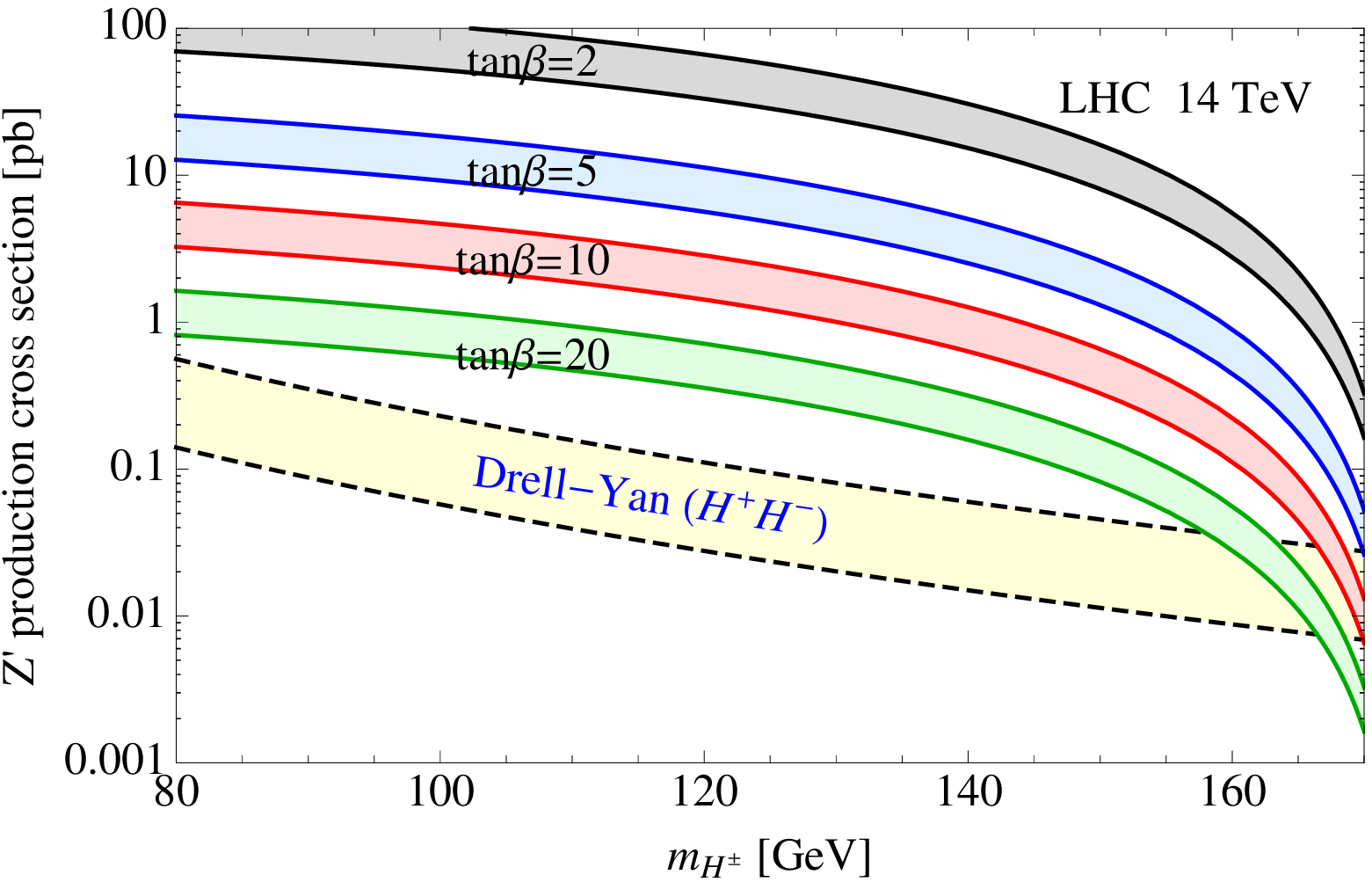}
\end{center}
\vspace{-0.1cm}
\caption{
(left) $\br(t \to b H^+)$ for $\tan\beta = 2$, $5$, $10$, $20$.
The $t$ decay into $H^+$ is larger for a smaller $m_{H^\pm}$ and smaller $\tan\beta$.
Production cross sections of the $Z'$ with $m_{H^\pm}$ in $t \bar t$ channel ($p p \to t \bar t \to b W \bar b W + Z'$) in (middle) 8 TeV LHC and (right) 14 TeV LHC.
The 14 TeV LHC provides about 4 times of the 8 TeV LHC cross section.
In the $m_W \lesssim m_{H^\pm} \lesssim m_t$ range, the results for $\tan\beta = 2$, $5$, $10$, $20$ are shown.
Drell-Yan channel ($p p \to H^+ H^- \to W W + Z' Z'$) (Dashed) is also shown for comparison.
The band indicates $\br (H^\pm \to W Z') = 0.5 - 1$ range.
\label{fig:comparison}}
\end{figure}

The branching fraction of top quark into charged Higgs and $b$ is shown in the left panel of Fig. \ref{fig:comparison} as a function of charged Higgs mass. 
A higher branching fraction is obtained for a lower $\tan\beta$.
Current limit allows about {\cal O}(1)\% branching fraction.
Top pair production and its decay to $Z^\prime$ provides a main production mechanism of $Z^\prime$, {\it i.e.,}
$\sigma(p p \to b W^+ \, \bar b W^- + Z'\text{s}) \simeq \sigma_{t \bar t} \, 2 X $, 
where $X \equiv BR( t \to b W + Z^\prime s) = BR(t \to b H^+ ) BR( H^\pm \to W^\pm Z^\prime) $.
We will take a rather conservative value of $X$ = 0.001, 
and assume $BR( Z^\prime \to \ell^+\ell^-)=0.2$, for our illustration, meaning 0.02\% of top quark decay is into the lepton pairs.
In Fig. \ref{fig:comparison}, production cross sections of $Z^\prime$ are shown as a function of charges Higgs mass for a few choices of $\tan\beta$ 
at the 8 TeV (middle) and 14 TeV (right).
Drell-Yan production has negligible model dependence and its cross sections are shown in yellow band with dashed curve.
$Z^\prime$ production via top quark is shown in solid band for $\tan\beta=$ 2, 5, 10 and 20.
The band indicates 50-100\% of branching fraction of the charged Higgs into $W Z^\prime$.
Cross section at 14 TeV is about 4 times larger than that at 8 TeV.
For relatively lower values of $\tan\beta$, top pair production is more important for $Z^\prime$ production.

\section{Current bounds and prospect at the 14 TeV LHC}

\begin{figure*}[t] 
\begin{center}
\includegraphics[width=0.42\textwidth,clip]{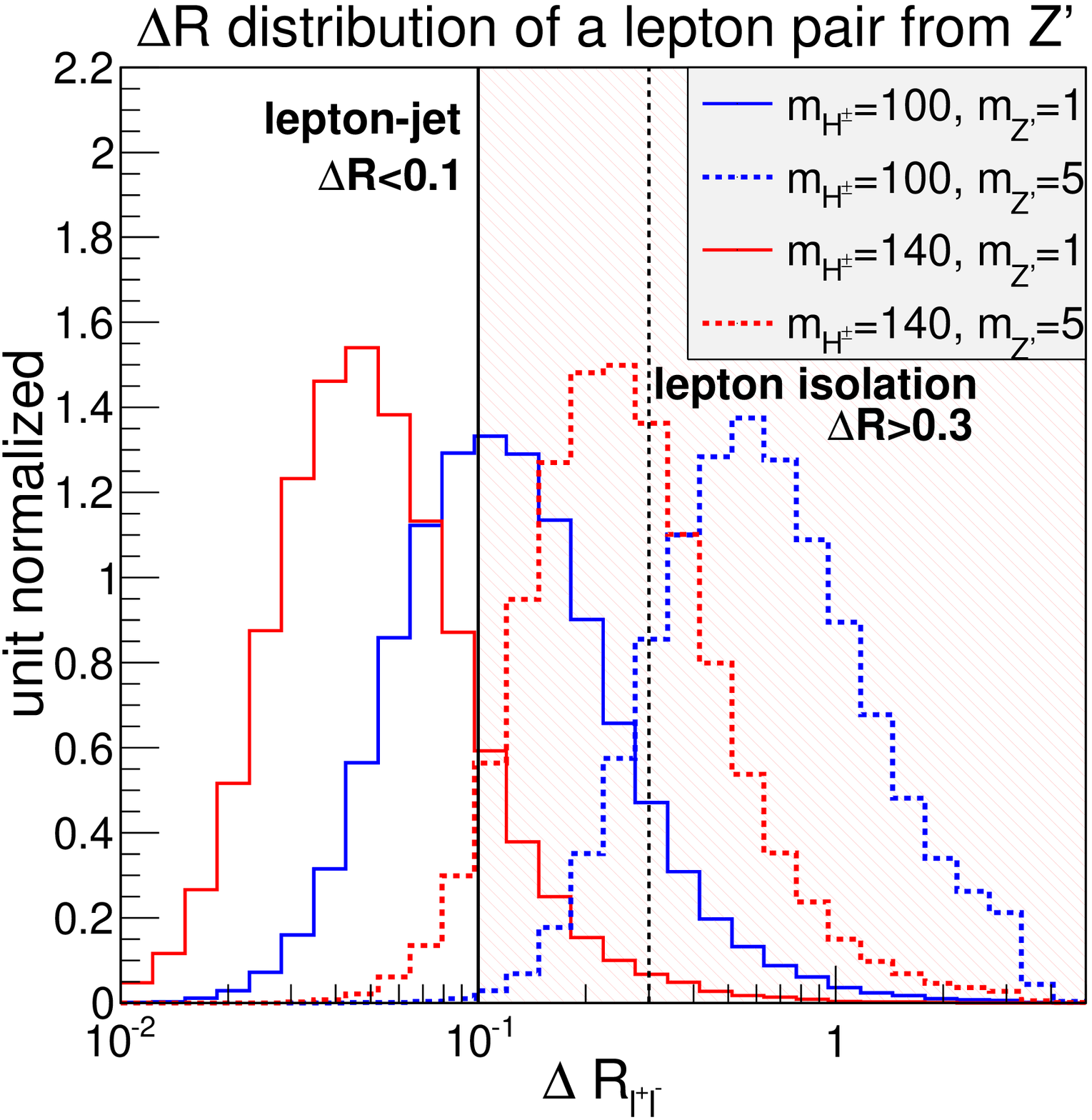} \hspace{0.5cm}
\includegraphics[width=0.42\textwidth,clip]{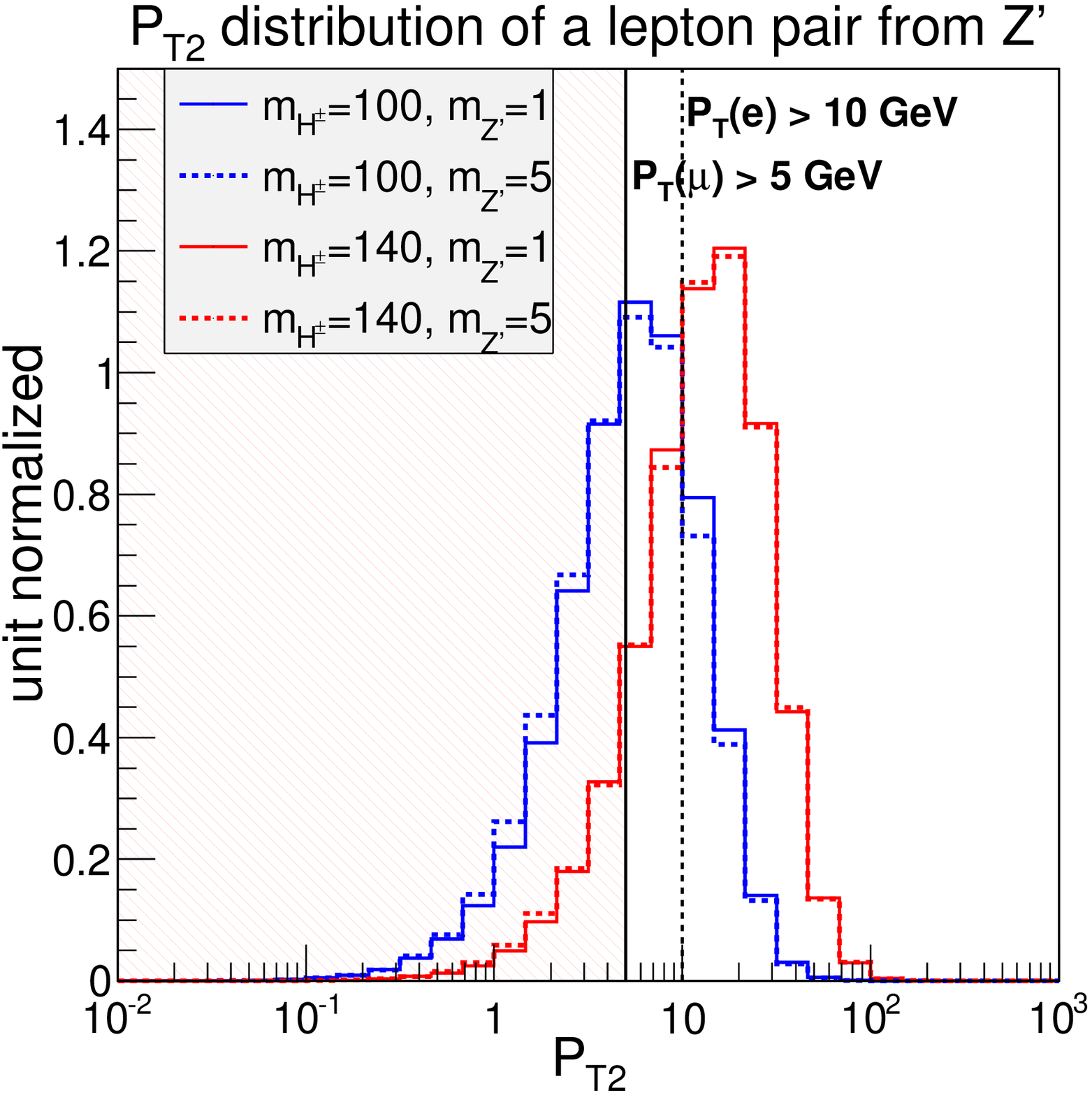}
\caption{\label{fig:iso} For $14 ~\tev$ LHC, we plot parton level lepton information for $\Delta R$ of lepton pair from $Z'$ in (left) and $P_{T2}=\min[{P_{T(\ell)}, P_{T(\bar\ell)}]}$ in (right)
to understand lepton-jet tagging efficiency.} 
\label{fig:why}
\end{center}
\end{figure*}

Once $Z^\prime$ is produced, it decays to leptons with significant branching fraction.
However, the light $Z^\prime$ can not be reconstructed with the usual tagging due to the following reason.
Invariant mass of the lepton pair can be expressed as
\bea
m_{\ell^+\ell^-}^2 = 2 P_{T_1} P_{T_2} \left(\cosh{\Delta \eta}-1 \right) 
\simeq 2 P_{T_1} P_{T_2} \left(\cosh{\Delta R}-1 \right) \, ,
\label{eq:mll}
\eea
with observation $\Delta R \simeq \Delta \eta$ since $\Delta \phi$ distribution is peaked at 0.
For a moderate lepton tagging efficiency, most analyses require a minimum $P_T$ cut, 
\begin{equation}
P_{T(e)}^{\textrm{min}} = 10 ~\gev \quad  {\rm and}  \quad P_{T(\mu)}^{\textrm{min}} = 5 ~\gev .
\end{equation}
Now an isolation requirement of $\Delta R > 0.3$ leads to the minimum invariant mass of 3 GeV for dielectron and 1.5 GeV for dimuon. 
Figure \ref{fig:why} shows distributions of $\Delta R$ and $P_t$ for two different masses of charged Higgs and $Z^\prime$.
$\Delta R = 0.1$, $0.3$ and $P_T = 5$, $10$ GeV lines are also drawn for comparison. 
As shown in the figure, most signals are cut off with the usual cuts.
Therefore conventional analysis would miss $Z^\prime$ lighter than these values.
We adopt an analysis method with the ``lepton-jet'' (LJ) proposed in Ref.~\cite{ArkaniHamed:2008qp}. 
Since we consider $t \bar t + \ell^+ \ell^-$ (from $\gamma^* / Z^*$) as a major irreducible background, 
we follow the LJ definition in Ref.~\cite{Cheung:2009su} as follows.
\begin{itemize}
\item[1.] At least two same flavor leptons with $P_T > 10 ~\gev$ (electron), $5 ~\gev$ (muon)  and in a cone of $\Delta R<0.1$.
\item[2.] Isolation: Hadronic and leptonic isolation of $\sum P_T < 3 ~\gev$ in $0.1<\Delta R<0.4$. 
\end{itemize}
Additionally, to suppress the backgrounds, we require $20\%$ window of the expected $Z'$ mass for an invariant mass of lepton-jet.
\begin{itemize}
\item[3.] Invariant mass cut on lepton-jet: $|m_{\textrm{LJ}} -m_{Z'}| \le 0.2\times m_{Z'}$.  
\end{itemize}

For a Monte Carlo simulation, we add $Z'$ and $H^\pm$ to the SM using FeynRules v2\,\cite{Alloul:2013bka} and simulate events with Madgraph v5.14\,\cite{Alwall:2011uj}, Pythia 6\,\cite{Sjostrand:2006za} and Delphes 3.0.11\,\cite{deFavereau:2013fsa}. We modified $b$-jet tagging/mis-tagging efficiency (tagging efficiency is around $60 - 75\%$ depending on $P_T$ and $\eta$) according to CMS CSVM tagging \cite{Chatrchyan:2012jua}.
For the lepton-jet analysis, we add a lepton-jet class to Delphes detector simulation.

We consider three signal boxes (dilepton channel, semi-lepton channel, and hadronic channel) with slight modification of CMS analyses according to number of triggered leptons. Jets are reconstructed with anti-$k_T$ algorithm with $\Delta R=0.5$. 
We require one $b$-tagged jet (or two $b$-tagged jets) and lepton-jet mass window as  we explained above.
Other improved event selection criteria are dependent on each channel as described in Ref.~\cite{Kong:2014jwa}.
We consider irreducible background only $(t \bar t + \ell^+ \ell^-)$ from virtual photon and virtual $Z$ boson radiations.
We do not consider possible reducible backgrounds from mistagged LJ.
For example, jets can be misidentified as electrons and overlapped leptons. 


We first obtain LJ tagging efficiencies at both 8 TeV and 14 TeV for several masses of charged Higgs and $Z^\prime$, as shown in Table \ref{table:effS8}. The numbers in parenthesis are efficiencies, 
when we require additional selection cuts following CMS analysis.
Then with cross section in Fig. \ref{fig:comparison} we can easily compute the number of signal events. 
The left panel in Table \ref{table:events} summarizes the number of signal and backgrounds events at 8 TeV LHC with 20 fb$^{-1}$.
At 8 TeV, $t\bar t$ production cross section is about 239 pb.
Then the number of signal events is calculated by product of cross section, branching fraction, efficiency and luminosity.
For example, the charged Higgs of 140 GeV and $Z^\prime$ of 2 GeV gives about 60 events over 5 background events.
On the other hand, conventional cuts give 4 signal events, which will be buried in the background uncertainty around 591, 
where the number of background events is $1.7 \times 10^4$.
Finally the right panel in Table \ref{table:events} shows  the required luminosity for $5 \sigma$ discovery at 14 TeV LHC.

\begin{table*}[t]
\begin{center}
{\small
    \tabcolsep 3.7pt
\centering
\begin{tabular}{|c|c||c|c|c||c|c|c|} 
\hline
LHC &$m_{Z'}$& \multicolumn{3}{c||}{$m_{H^\pm}$} & Mass range of & $\sigma_\text{LO}$  & $\epsilon_\textrm{LJ} (\epsilon_\textrm{(LJ+CMS)})$\\
\hhline{~~---}
[$\tev$] & [$\gev$]& $100 ~\gev$ & $140 ~\gev$ & $160 ~\gev$ & $ m_{\ell^+ \ell^-}\, [\gev]$ & [pb] & [\%]  \\
\hhline{--------}
\multirow{ 3}{*}{$8$ } & $1$ & 16.39 (4.30/2.09)  &  46.95 (11.08/4.41) & 51.99 (9.63/3.02)&	$(0.5 - 1.5)$ & $0.617$ & 1.86 (0.56/0.27)    \\
& $2$ &  3.11 (0.96/0.49) &  31.17 (7.71/3.11)  &   40.85 (7.58/2.55) &	$(1.0 - 3.0)$  & $0.157$  & 0.55 (0.20/0.11)  \\
& $5$  & 0.02 (0.00/0.00)  & 2.23 (0.65/0.29)  &   5.55 (1.22/0.42) &	$(3.0 - 5.0)$  & $0.0175$    & 0.29 (0.09/0.04) \\
\hhline{--------}\hhline{--------}
\multirow{ 3}{*}{$14$ } &$1$ & 16.10 (4.10/1.96)  &  44.53 (10.95/4.50) & 50.69 (9.54/3.15)  &	$(0.5 - 1.5)$ & $2.536$ & 1.82 (0.51/0.22)    \\
& $2$ &  3.32 (1.12/0.53) &  29.86 (7.53/3.14)  &   39.17 (7.70/2.51)  &	$(1.0 - 3.0)$ & $0.640$  & 0.53 (0.21/0.10)  \\
& $5$  & 0.03 (0.01/0.00)  & 2.54 (0.68/0.31)  &   5.92 (1.36/0.47)  &	$(3.0 - 5.0)$ & $0.0706$    & 0.38 (0.14/0.07) \\
\hline
\end{tabular}
\caption{Lepton-jet tagging efficiency $\epsilon_\textrm{LJ}$ (in $\%$), for given $m_{H^\pm}$ and $m_{Z'}$, for $p p \to b W \bar b W + \ell^+ \ell^-$ from virtual photon and virtual $Z$ boson in $8\,\tev$ and $14\,\tev$ LHC (and efficiency in parenthesis if we require additional selection cuts, 
$\epsilon_\textrm{(LJ+CMS[1b])} / \epsilon_\textrm{(LJ+CMS[2b])}$ with requiring one $b$-tagged or two $b$-tagged jets.)
Coupling structure of $Z'$ to the lepton does not give any significant effect on the tagging efficiency. In the above table, we take axial coupling as an example. 
For backgrounds, we set the trigger of $m_{\ell^+ \ell^-}$ mass window as in table to enlarge statistics.
\label{table:effS8}
 }}
 \end{center}
\end{table*}

\begin{table}[h]
\centering
\begin{tabular}{ll}
\begin{tabular}{|c||c|c|c|c|} 
\hline
$m_{Z'}$ & \multicolumn{3}{c|}{$m_{H^\pm}$} & \\
\hhline{~---~}
[GeV]& $100 ~\gev$ & $140 ~\gev$ & $160 ~\gev$ & BKG\\
\hhline{-----}
$1$ & 40.0 & 84.3  & 57.7  &  66.6 \\
$2$ & 9.4 & 59.5  & 48.8  & 6.9  \\
$5$ & 0.0 & 5.5 & 8.0  & 0.3   \\ 
\hline
\end{tabular}
&
\hspace{0.5cm}
\begin{tabular}{|c||c|c|c|} 
\hline
 $m_{Z'}$ & \multicolumn{3}{c|}{$m_{H^\pm}$}  \\
\hhline{~---}
[GeV]& $100 ~\gev$ & $140 ~\gev$ & $160 ~\gev$\\
\hhline{----}
$1$ & $6.3\,\fb^{-1}$ & $1.4\,\fb^{-1}$  & $2.7\,\fb^{-1}$ \\
$2$ & $11.7\,\fb^{-1}$ & $0.7\,\fb^{-1}$ & $1.0\,\fb^{-1}$  \\
$5$ & - & $6.2\,\fb^{-1}$  & $3.4\,\fb^{-1}$   \\ 
\hline
\end{tabular}
 \end{tabular}
 \caption{(left) Expected number of events in each lepton-jet bin ($20\%$ window) with two $b$-tagging at the $8 ~\tev$ LHC with $20 ~\fb^{-1}$. We set $X=0.001$ and $\textrm{BR}(Z'\to\ell^-\ell^+)=0.2$.
Signal events were obtained with high order $\sigma_{t \bar t}$ with branching ratio, and
the background events were obtained with tree-level simulation with $K_\text{bkg} = 2$.
(right) Required luminosity at the $14 ~\tev$ LHC for the likelihood ratio $S_\textrm{cL} = 5$.
\label{table:events}
}
\end{table}

 \section{Summary}

 A light $Z^\prime$ of roughly O(1) GeV with small coupling is a very well motivated new physics candidate as it can address some astrophysical anomalies as well as the muon $g$-2 anomaly. 
 While its search is very active at the low energy experimental facilities, its search at the LHC is relatively limited so far.
In this paper, we discussed prospect of such a light $Z^\prime$ gauge boson through a top quark at the LHC. 
In particular, we considered a scenario in which the top quark decays through a charged Higgs, which in turn decays into $Z^\prime$ dominantly.
Even a small branching fraction of $t \to b + H^+$ can be enough to produce the $Z^\prime$ at the observable level at the LHC experiments.
For some parameter space, even the existing 8 TeV LHC data may give enough signals for a discovery. 
It also guarantees a huge discovery potential even at the very early stage of the 14 TeV LHC experiments.

Because of lightness of the $Z^\prime$, its decay products, the lepton pairs, are highly collimated forming lepton-jets and a careful definition of leptons is necessary. 
 As the top quark decay into dark sector mode can be easily mistaken with its dominant $bW$ mode, reanalysis of the top data at the hadron collider can possibly reveal interesting hint of the $Z^\prime$ even if the $Z^\prime$ is very elusive or decays invisibly. 
 It calls for attention in both experimental and theoretical sides of the top quark study.

\Acknowledgements
KK is supported by the U.S. DOE under Grant No.~DE-FG02-12ER41809, and the University of Kansas General Research Fund allocation 2301566.
HL is in part supported by U.S. DOE under Grant No.~DE-AC05-06OR23177, and the NSF under Grant No.~PHY-1068008.
MP is supported by World Premier International Research Center Initiative (WPI Initiative), MEXT, Japan.

\end{document}